\documentstyle[12pt,epsf]{article}
\newcommand{\rmt}{\rm\textstyle}

\def\be{\begin{equation}}
\def\ee{\end{equation}}
\def\bea{\begin{eqnarray}}
\def\eea{\end{eqnarray}}
\def\beas{\begin{eqnarray*}}
\def\eeas{\end{eqnarray*}}
\def\err#1#2#3  {{\it Erratum} {\bf#1} (#2) #3 }
\def\ib#1#2#3   {{\it ibid.} {\bf#1} (#2) #3 }
\def\ijmp#1#2#3 {{\em Int. J. Mod. Phys.} {\bf#1} (#2) #3 }
\def\jetp#1#2#3 {{\em JETP Lett.} {\bf#1} (#2) #3 }
\def\mpl#1#2#3  {{\em Mod. Phys. Lett.} {\bf#1} (#2) #3 }
\def\nat#1#2#3  {{\em Nature (London)} {\bf#1} (#2) #3 }
\def\nc#1#2#3   {{\em Nuovo Cim.} {\bf#1} (#2) #3 }
\def\nim#1#2#3  {{\em Nucl. Instr. Meth.} {\bf#1} (#2) #3 }
\def\np#1#2#3   {{\em Nucl. Phys.} {\bf#1} (#2) #3 }
\def\pcps#1#2#3 {{\em Proc. Cam. Phil. Soc.} {\bf#1} (#2) #3 }
\def\pl#1#2#3   {{\em Phys. Lett.} {\bf#1} (#2) #3 }
\def\prep#1#2#3 {{\em Phys. Rep.} {\bf#1} (#2) #3 }
\def\prev#1#2#3 {{\em Phys. Rev.} {\bf#1} (#2) #3 }
\def\prl#1#2#3  {{\em Phys. Rev. Lett.} {\bf#1} (#2) #3 }
\def\prs#1#2#3  {{\em Proc. Roy. Soc.} {\bf#1} (#2) #3 }
\def\ptp#1#2#3  {{\em Prog. Th. Phys.} {\bf#1} (#2) #3 }
\def\rmp#1#2#3  {{\em Rev. Mod. Phys.} {\bf#1} (#2) #3 }
\def\rpp#1#2#3  {{\em Rep. Prog. Phys.} {\bf#1} (#2) #3 }
\def\sjnp#1#2#3 {{\em Sov. J. Nucl. Phys.} {\bf#1} (#2) #3 }
\def\spj#1#2#3  {{\em Sov. Phys. JEPT} {\bf#1} (#2) #3 }
\def\zp#1#2#3   {{\em Zeit. Phys.} {\bf#1} (#2) #3 }
\begin{document}
\newcommand{\linespace}[1]{\protect\renewcommand{\baselinestretch}{#1}
  \footnotesize\normalsize}
\begin{flushright} 
FERMILAB Conf-96/228
\end{flushright}
\begin{center} 
\vspace*{4.cm} 
COMMENT ON LEPTOPHOBIC BOSONS AND $\nu N$ NEUTRAL CURRENT
SCATTERING DATA
\end{center} 
\vspace{5cm}
\begin{center} 
Kevin S.~McFarland, Fermilab
\end{center} 
\vspace{5cm} 
\begin{abstract} 
The relevance of existing $\nu N$ deep inelastic scattering
data to the model of an additional leptophobic vector boson
presented at this conference is discussed.  It is shown that
the neutral current data is in good agreement with the Standard Model
and disfavors such a leptophobic boson.
\end{abstract} 

\pagebreak

A number of different new physics models have been suggested to explain the
significant discrepancies between the measured $R_b$ and $R_c$ at LEP
and their Standard Model predictions.  Generally, these models fall into
two classes: those that affect primarily the third generation (not affecting
$R_c$), and those that introduce changes in couplings to all generations
of fermions.  The latter type of model takes advantage of the observation
that the quantity $3\delta\Gamma_b+2\delta\Gamma_c$ is consistent with zero,
where $\delta\Gamma_q$
is the difference between the Standard Model and measured widths of
$Z\to q\overline{q}$.  As has been pointed out by a number of
authors \cite{Chiappet}\cite{Altar}\cite{Kolda}, this suggests 
generation-universal shifts in the hadronic couplings 
to the $Z^0$ so that
\bea
\Gamma_{d,s,b}&=&\Gamma_{d,s,b}^{SM}+\delta\Gamma_b \\
\Gamma_{u,c,t}&=&\Gamma_{u,c,t}^{SM}+\delta\Gamma_c. 
\eea
Such a model not only shifts observables at $Z^0$ pole away from their
Standard Model values, but will also affect lower-energy neutral-current
phenomena such as atomic parity violation and $\nu N$ deep inelastic
scattering, both of which measure the interactions of first
generation quarks.

A model presented at the XXXI$^{\rmt i\grave{e}me}$ Rencontres de Moriond
Electroweak session \cite{Altar}\cite{DiB} introduced a new neutral
vector boson, $V$, with a $1$~TeV mass.  In order to avoid conflict
with very precise leptonic data from SLD and LEP, the coupling of the $V$
to leptons is chosen to be zero; thus the name ``leptophobic''.
However, such a boson will still affect processes induced by leptons,
be they neutrinos scattering from quarks or electrons annihilating to $Z^0$s
at resonance, as long as it has non-zero mixing with the Standard Model $Z^0$.
The mixing, parameterized by the angle $\xi$,  will change the couplings 
of the physical $Z^0$ to quarks,
\be
\epsilon_{L,R}~\to~\epsilon^{SM}_{L,R}\cos\xi+\epsilon^{V}_{L,R}\sin\xi,
\ee
and will add a tree level shift to the $\rho$ parameter,
\be
\rho~\to~\rho^{SM}+\xi^2\left( \frac{M_V}{M_Z}\right) ^2.
\ee
The $V^0$ couplings in this model are given by $\epsilon^{u,d}_L=x$,
 $\epsilon^u_R=y_u$ and $\epsilon^d_R=y_d$.

\begin{table}[b]
\begin{tabular}{|r|c|c|} \hline
& Fit 1 & Fit 2 \\ \hline
 $\xi$ & $2.8\times10^{-3}$ & $3.8\times10^{-3}$ \\
 $x$ & -1.8 & -1.0 \\
 $y_u$ & 4.7 & 2.7 \\ \hline
 $\chi^2$ & 14.3/9 dof & 15.9/9 dof \\ \hline
\end{tabular}
\begin{tabular}{|r|c|c|c|c|} \hline
& Measured Value & Std.\ Model & Fit 1 & Fit 2 \\ \hline
 $g_L^2$ & $0.3017\pm0.0033$ & $0.303$ & $0.3045$ & $0.3049$ \\
 $g_R^2$ & $0.0326\pm0.0033$ & $0.030$ & $0.0266$ & $0.0271$ \\ \hline
 $\chi^2$ & & 0.77/2 dof & 4.03/2 dof & 3.72/2 dof \\ 
 C.L. & & 68\% & 13\% & 15\% \\ \hline
\end{tabular}
\caption{The reported best fit values for the parameters of the 
leptophobic boson model are shown in the left table.  The $\chi^2$
values do not include the neutrino data. 
The right table shows the $\nu N$ constraints on neutral current 
quark couplings, 
  and the effects of the leptophobic boson models.  
  Only the isoscalar combinations of
  couplings $g_{L,R}^2=\left( \epsilon^u_{L,R}\right) ^2+\left( 
      \epsilon^d_{L,R}\right) ^2$
  are measured to high-precision in $\nu N$ scattering.}
\label{tab:nuN}
\end{table}

In the presentation at Moriond \cite{DiB}, the couplings $x$, $y_u$
and the mixing angle, $\xi$ were allowed to float in a fit to $12$
observables, $\Gamma_Z$, $R_l$, $\sigma_h$, $R_b$, $R_c$, $M_W/M_Z$,
 ${\cal A}_l$, ${\cal A}_b$, ${\cal A}_c$, $A^b_{FB}$, $A^c_{FB}$ and
the weak charge in Cesium, $Q_W$.  $M_V$ was fixed at 1~TeV, and $y_d$
was set to zero.  Two of the reported fit results are shown in
Table~\ref{tab:nuN}.  The second set of fit parameters shown
are claimed to be in better agreement with the extremely high $p_T$
CDF jet rates than the first \cite{Altar}\cite{DiB}.

\begin{figure}[t]
\epsfxsize=\textwidth
\epsfbox{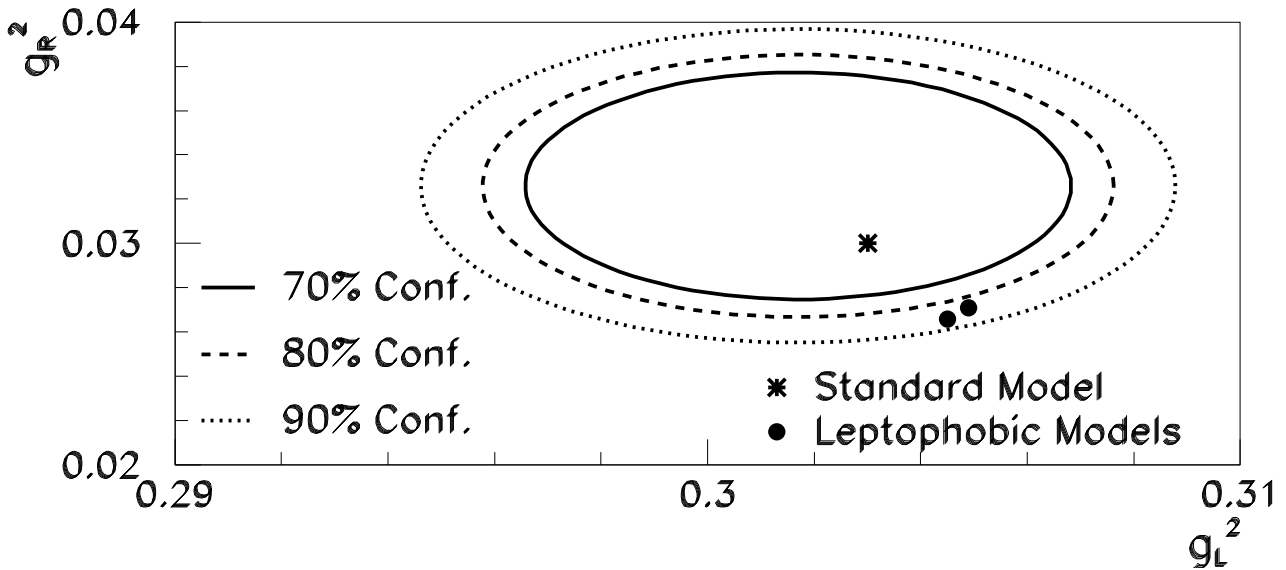}
\caption{The $\nu N$ data and leptophobic models in the $g_L^2$-$g_R^2$ plane}
\label{fig:nuNcl}
\end{figure}

Global fits to the neutral-current quark couplings from $\nu N$
scattering data have been performed \cite{lang}\cite{foglihaidt},
and recently the CCFR collaboration has reported their results in
terms of quark couplings \cite{oldCCFR}\cite{newCCFR}.  The constraint
on the neutral current quark couplings from the $\nu N$ data used for
this analysis is that reported in the Review of Particle Properties
\cite{PDG} which does not reflect the recent update to the
CCFR data \cite{newCCFR}.  Table~\ref{tab:nuN} shows the measured values,
Standard Model predictions, and the values predicted from the two fits 
presented above.  The $\nu N$ data is in agreement with the Standard
Model, but excludes the central values of either leptophobic fit at about 
85\% confidence.  This is shown graphically in Figure~\ref{fig:nuNcl}.

In conclusion, models which introduce generation-universal changes to
neutral current quark couplings to explain the $R_b$, $R_c$ at LEP will
also induce changes in parameters measured accurately in $\nu N$ scattering.
Although the $\nu N$ does not rule out models such as the one discussed
here, it clearly can place constraints on such models.  For example,
in the preferred fit (\#2) of this leptophobic boson model \cite{DiB},
the $\chi^2$, including the $\nu N$ data, is now 19.6/11 dof, which is
an improvement over the Standard Model 29.0/14 dof, but is still excluded
at more than $95\%$ confidence.

The author wishes to thank Alain Blondel for the initial suggestion
that the neutrino data might constrain these models, Nicola Di
Bartolomeo for an enjoyable public debate on the subject, and Paul
Langacker and Chris Kolda for sharing their expertise on this subject.


\vspace{-.15in} 
\begin{thebibliography}{99}
\small
\linespace{1.0} 
\vspace{-10pt}
\bibitem{Chiappet} P.~Chiappetta {\em et al.\,}, PM/96-05, hep-ph/9601306.\vspace{-10pt}
\bibitem{Altar} G.~Altarelli {\em et al.\,}, CERN-TH/96-20, hep-ph/9601324.\vspace{-10pt}
\bibitem{Kolda} K.S.~Babu {\em et al.\,}, IASSNS-HEP-96/20, hep-ph/9603212.\vspace{-10pt}
\bibitem{DiB} N.~DiBartolomeo, proceedings of the XXXI$^{\rmt i\grave{e}me}$ 
Recontres de Moriond, March 1996.\vspace{-10pt}
\bibitem{lang} U.~Amaldi {\em et al.\,}, \prev{D36}{1987}{1385}.\vspace{-10pt}
\bibitem{foglihaidt} G.~Fogli, D.~Haidt, \zp{C40}{1988}{379}.\vspace{-10pt}
\bibitem{oldCCFR} C.~Arroyo, B.J.~King {\em et al.\,}, \prl{72}{1994}{3452}.\vspace{-10pt}
\bibitem{newCCFR} K.S.~McFarland {\em et al.\,}, 
 proceedings of the XXXI$^{\rmt i\grave{e}me}$ 
Recontres de Moriond, March 1996, FERMILAB-Conf-96/227.\vspace{-10pt}
\bibitem{PDG} L.~Montanet {\em et al.\,}, Review of Particle Properties,
\prev{D50}{1994}{1173}.\\
 J.~Erler, P.~Langacker, \prev{D52}{1995}{441}.\vspace{-10pt}
\end{thebibliography}
\end{document}